\documentclass[12pt]{iopart}

\usepackage{hyperref}
\hypersetup{
colorlinks=true,
linkcolor={blue},
citecolor={blue},
urlcolor={blue},
bookmarksnumbered=true,
bookmarksopen=true,
breaklinks=true, 
hypertexnames=false, 
}

\usepackage{graphicx} 
\usepackage{braket} 

\newcommand{\mtext}[1]{\mathrm{#1}} 

\begin{document}

\title[SWAP gate realized with CZ and iSWAP gates in a superconducting architecture]{Quantum SWAP gate realized with CZ and iSWAP gates in a superconducting architecture}

\author{Christian Križan$^1$, Janka Biznárová$^1$, Liangyu Chen$^1$, \\ Emil Hogedal$^1$, Amr Osman$^1$, Christopher W. Warren$^1$, Sandoko Kosen$^1$\footnote{Present address: Nord Quantique, 1950 rue Roy, Sherbrooke, QC J1K 1B7, Canada}, Hang-Xi Li$^1$, Tahereh Abad$^1$, \\ Anuj Aggarwal$^1$, Marco Caputo$^2$, Jorge Fernández-Pendás$^1$, Akshay Gaikwad$^1$, Leif Grönberg$^2$, Andreas Nylander$^1$, \\ Robert Rehammar$^1$, Marcus Rommel$^1$, Olga I. Yuzephovich$^1$, Anton Frisk Kockum$^1$, Joonas Govenius$^2$, Giovanna Tancredi$^1$, and Jonas Bylander$^1$}

\address{$^1$ Department of Microtechnology and Nanoscience, Chalmers University of Technology, SE-412 96 Gothenburg, Sweden}
\address{$^2$ VTT Technical Research Centre of Finland, Ltd., QTF Centre of Excellence, FI-02044 VTT Espoo, Finland}
\ead{krizan@chalmers.se, christian.krizan@gmail.com}
\vspace{10pt}
\begin{indented}
\item[]December 2024
\end{indented}

\begin{abstract}
It is advantageous for any quantum processor to support different classes of two-qubit quantum logic gates when compiling quantum circuits, a property that is typically not seen with existing platforms. In particular, access to a gate set that includes support for the CZ-type, the iSWAP-type, and the SWAP-type families of gates, renders conversions between these gate families unnecessary during compilation as any two-qubit Clifford gate can be executed using at most one two-qubit gate from this set, plus additional single-qubit gates. We experimentally demonstrate that a SWAP gate can be decomposed into one iSWAP gate followed by one CZ gate, affirming a more efficient compilation strategy over the conventional approach that relies on three iSWAP or three CZ gates to replace a SWAP gate. Our implementation makes use of a superconducting quantum processor design based on fixed-frequency transmon qubits coupled together by a parametrically modulated tunable transmon coupler, extending this platform's native gate set so that any two-qubit Clifford unitary matrix can be realized using no more than two two-qubit gates and single-qubit gates.
\end{abstract}
%
\vspace{2pc}

\section{Introduction} \label{sec:introduction}

The available gate set is an important consideration for practical quantum computation on the hardware platform of choice. A complete single-qubit gate set, supplemented with an entangling two-qubit operation, is a sufficient resource for universal quantum computation \cite{Kitaev1999, Barenco1995}.

In practice, however, different types of two-qubit gates can be advantageous depending on the particularities of the hardware and its circuit architecture \cite{Leymann2020}, such as the feasible qubit connectivity \cite{Leymann2020, Abrams2020, McEwen2023, Yan2024}, the presence of residual interactions in the Hamiltonian leading to coherent errors, and the dominant noise channels leading to incoherent errors \cite{Iverson2020, Abad2022, Abad2023}. 
For superconducting qubits, the circuit-QED architecture offers flexibility in addressing such design trade-offs for quantum processors \cite{Blais2021}.

\begin{figure}[t!]
    \centering
    \includegraphics[width=0.50\textwidth]{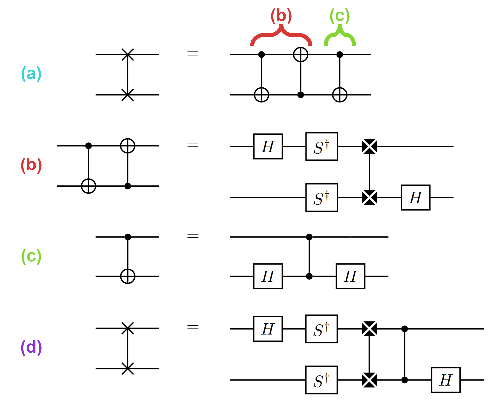}
    \caption{ \scriptsize Derivation of the SWAP operation using CZ and iSWAP gates. 
    (a) The SWAP gate is equivalent to three subsequent CNOT gates \cite{Mike_and_Ike}, as is true also in classical logic. 
    (b) The initial double-CNOT gate may be expressed using an iSWAP gate combined with conjugate-transposed phase gates ($S^\dagger$) and Hadamard ($H$) gates \cite{Crooks2023}: $\mtext{CNOT}_{10}(\mtext{CNOT}_{01}(\Psi))$ is decompiled into its iSWAP equivalent. 
    (c) We express the last $\mtext{CNOT}_{01}$ gate from (a) as its local equivalent \cite{Zhang2004}, using a CZ gate and by straddling one qubit with two $H$ gates. 
    (d) We finally put (b) and (c) together to form a SWAP operation. Note here that the two subsequent $H$ gates cancel into identity, $H(H(\Psi)) = I(\Psi)$. In principle, the order of the CZ and iSWAP gates in (d) is irrelevant as they commute.}
    \label{fig:SWAP_derivation}
\end{figure}

When executing quantum algorithms, it is often desirable to have easy access to a larger gate set than the bare minimum to enable efficient quantum circuit compilation \cite{Leymann2020} even though combinations of two-qubit and single-qubit gates can, in principle, perform conversions between two-qubit gate families \cite{McEwen2023, Zhang2003, Crooks2023}.
The Clifford group contains four such families of locally inequivalent two-qubit gates \cite{Crooks2023}: CNOT-like gates such as the CZ gate, iSWAP-like gates, SWAP-like gates, and identity \cite{Zhang2003, Sung2021, Lao2022, Huang2023}. Gates are said to be locally inequivalent when one cannot be converted into another using only single-qubit rotations, which typically feature higher fidelity than two-qubit operations and execute $\sim$10x faster.
An example of a conversion between a CNOT gate and an iSWAP gate is shown in figure~\ref{fig:SWAP_derivation}.

In devices made for noisy intermediate-scale quantum (NISQ)~\cite{Preskill2018} computation, quantum circuits run within their constituent qubits' coherence time and without error correction. 
Here for example, the CZ gate is advantageous in some NISQ algorithms based on minimizing the energy of the Ising Hamiltonian~\cite{Bengtsson2020}, which includes an oftentimes residual and always-on ZZ interaction that commutes with the CZ unitary matrix.

In another example, the iSWAP family is naturally suited for quantum chemistry calculations where the qubits represent fermionic states; here, iSWAP plus single-qubit gates form the fSWAP interaction, which respects fermionic anti-commutation relations. This interaction has known decompositions using four CNOT-like gates \cite{CerveraLierta2018}, although two CNOT gates is theoretically sufficient \cite{Williams2011} as $\mtext{fSWAP}\in SO(4)$.

However, iSWAP gates do not commute with ZZ terms \cite{Ganzhorn2020}, which is why errors of such always-on interactions may accumulate. Even though the CZ gate is implemented via a Hamiltonian that also does not commute with ZZ terms, this gate may have its pulse parameters adjusted to omit static ZZ terms \cite[Section \uppercase\expandafter{\romannumeral5\relax}.D]{Ganzhorn2020}.

In the context of fault-tolerant quantum computation, proposals for subcells of the error-correcting surface code use the CNOT gate, locally equivalent to the CZ gate, as their fundamental two-qubit gate \cite{Fowler2012}. 
On the other hand, in surface-code circuits, the iSWAP gate may reduce the coupling per qubit and boosts resilience to leakage compared to CNOT-like gates \cite{McEwen2023}.

Superconducting qubits afford versatility in the implementation of two-qubit gates as we can engineer interactions between states, and have access to a rich level structure. One prominent implementation is the parametrically modulated coupler, which, with transmon qubits, has been used to demonstrate high-fidelity two-qubit gates \cite{Ganzhorn2020, Roth2017, McKay2016, Xu2020, Stehlik2021} and to run quantum algorithms \cite{Bengtsson2020}.
The device used in the investigation reported here comprises a frequency-tunable transmon coupler that is connected to two fixed-frequency transmon qubits, see figure~\ref{fig:chip_and_energy_levels}. Microwave modulation of the coupler's frequency mediates interactions between the qubits. In this system, we can address the different transitions in the two-qubit manifold, which enables native execution of CZ and iSWAP gates.

By means of microwave pulses, the CZ gate can be implemented by driving the $\ket{11} \rightarrow \ket{20}$ (or $\ket{11} \rightarrow \ket{02}$) transition; the iSWAP gate relies on XX+YY exchange interaction \cite{Roth2019}, activated by driving the $\ket{01} \rightarrow \ket{10}$ transition; and the bSWAP uses the XX-YY exchange interaction, which can be realized by driving the two-photon $\ket{00} \rightarrow \ket{11}$ transition \cite{Roth2017, Roth2019,Poletto2012}. We call these gates native to this system, whereas for example CNOT is not native, since we compose it from CZ and single-qubit gates, shown in figure~\ref{fig:SWAP_derivation}(c).

One important gate that is not native to most \cite{Leymann2020, Han2024} quantum processors is the SWAP gate. This gate is useful in systems without long-range or all-to-all couplings between qubits, which is typical for superconducting quantum processor architectures. The SWAP gate allows a two-qubit operation between two distant qubits to be performed by repeatedly swapping the states \cite{Abrams2020, Yan2024} within a chain of qubits, until the desired qubit states come adjacent to one another \cite{Yan2024}. Moreover, the SWAP-like family of gates is the only two-qubit gate type that is guaranteed to avoid creating entanglement for any separable input state \cite{Rezakhani2004}. Notably, the SWAP gate is native to quantum-dot-based quantum processors through isotropic Heisenberg exchange coupling \cite{Ni2023, Taylor2024}. Native support for the SWAP-like family of gates is rare in other platforms, but has been shown in recent advancements alongside support for iSWAP-like gates and the B gate, in a different superconducting device platform~\cite{Wei2024}. Whenever the SWAP gate is not natively supported, it can be realized by either three CZ gates \cite{Lukac2010, Sangouard2005}, three iSWAP gates \cite{Venturelli2018}, or three CNOT gates in sequence \cite[figure~\ref{fig:SWAP_derivation}]{Mike_and_Ike}.

In this paper, we show experimentally that we can implement SWAP operations more simply by combining one CZ gate and one iSWAP gate in sequence, with the addition of single-qubit gates. We note that CZ and iSWAP gates commute, so that they can be applied in any order---and even simultaneously, which would further reduce the total gate time, compared to the longer concatenation required by the three consecutive CZ, iSWAP, or CNOT gates.
Our work represents an improvement at the circuit compilation level; we augment the native gate set without changing the architecture. Our implementation reduces the length of compiled quantum algorithms produced by contemporary circuit synthesis tools where SWAP operations are realized as trivial solutions of three CNOT-like gates, or three iSWAP-like gates.
Such enhancements are particularly crucial considering NISQ-era gate-error constraints given the high number of consecutive gates that must be implemented during the course of an algorithm \cite{Leymann2020, Abrams2020, Yan2024, Preskill2018}. 
Besides NISQ, an augmented gate set may become useful in quantum error correction, where encoding and decoding segments can benefit from circuits designed to minimize the impact of SWAP gates~\cite{Mondal2024}.

\section{SWAP through combined CZ and iSWAP gates} \label{sec:SWAP_decomposition_derived}
This section derives the proposed SWAP operation, composed of an iSWAP gate followed by a CZ gate. The role of the iSWAP gate is to exchange state information, albeit with unwanted added phases for some input states, which are conditionally corrected by the CZ gate.

Our quantum circuits assume right-handed rotations about Bloch sphere axis $N \in [X,Y,Z]$, where $\ket{0}$ is located at the north pole. We denote \textit{increasing} a phase as a counter-clockwise Z-axis rotation of the Bloch vector when viewed from $\ket{0}$. The basis elements of our two-qubit computational states is given as $[\ket{00}, \ket{01}, \ket{10}, \ket{11}]^T$, and accordingly, we define our two-qubit gates as the following unitary operations:

\begin{equation}\label{eq:U_CZ}
U_{\mtext{CZ}} = 
\left(
\begin{array}{cccc}
1 & 0 & 0 & 0 \\
0 & 1 & 0 & 0 \\
0 & 0 & 1 & 0 \\
0 & 0 & 0 & -1
\end{array}
\right),
\end{equation}

\begin{equation}\label{eq:U_iSWAP}
U_{\mtext{iSWAP}} = 
\left(
\begin{array}{cccc}
1 & 0 & 0 & 0 \\
0 & 0 & i & 0 \\
0 & i & 0 & 0 \\
0 & 0 & 0 & 1
\end{array}
\right),
\end{equation}

\begin{equation}\label{eq:U_SWAP}
U_{\mtext{SWAP}} = 
\left(
\begin{array}{cccc}
1 & 0 & 0 & 0 \\
0 & 0 & 1 & 0 \\
0 & 1 & 0 & 0 \\
0 & 0 & 0 & 1
\end{array}
\right).
\end{equation}

We perform $U_{\mtext{SWAP}}$ using our equivalent quantum circuit derived in figure~\ref{fig:SWAP_derivation}. The quantum circuit in figure~\ref{fig:SWAP_derivation}(d) uses conjugate-transpose phase gates ($S^\dagger$) and Hadamard ($H$) gates, defined as

\begin{equation} \label{eq:U_s_dagger}
S^\dagger = 
\left(
\begin{array}{cc}
1 & 0\\
0 & -i
\end{array}
\right),
\end{equation}

\begin{equation}
H = \frac{1}{\sqrt{2}} 
\left(
\begin{array}{cc}
1 & 1\\
1 & -1
\end{array}
\right),
\end{equation}

\noindent where the $S^\dagger$ gates are implemented as virtual-Z gates, explained in Section \ref{section:two_qubit_gate_characterization}.

\section{Device}

\begin{figure}[t!]
    \centering
    \includegraphics[width=0.50\textwidth]{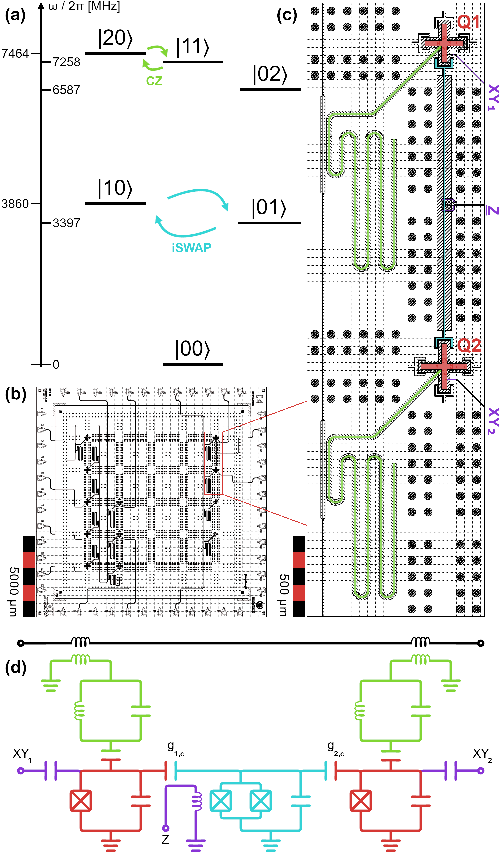}
    \caption{ \scriptsize (a) Scale drawing of the energy-level diagram for our two-qubit system $\ket{Q_1\,Q_2}$, where the coupler remains in the ground state for all experiments. We indicate the native two-qubit gate interactions used in our experiments. 
    (b) Two-tier flip chip quantum processor. 
    (c) Device used in our experiments: the qubits and coupler are placed on the flip chip die, and the resonators are located on a silicon interposer \cite{Kosen2022, Kosen2024}. The interposer contains the signal wiring that provides charge to the qubits, labeled XY$_{\mtext{1}}$ and XY$_{\mtext{2}}$, and flux to the coupler, labeled Z.
    (d) Equivalent circuit-QED schematic of (c), with resonators highlighted in green, transmons highlighted in red, the coupler highlighted in turqoise along with its coupling strengths g, and signal wiring hightlighted in purple. See \ref{appendix:experimental_setup} for the full setup.}
    \label{fig:chip_and_energy_levels}
\end{figure}

For the experimental demonstration, we used two fixed-frequency transmon qubits \cite{Koch2007} of a flip chip quantum processor shown in figure~\ref{fig:chip_and_energy_levels}, based on Chalmers' 3D architecture \cite{Kosen2022, Kosen2024}. 
We drive single-qubit state transitions using individual ``XY'' control lines and 
two-qubit transitions via a parametrically modulated, flux-tunable transmon
coupler \cite{McKay2016} using a ``Z'' control line.

\begin{figure}[t!]
    \centering
    \includegraphics[width=1.00\textwidth]{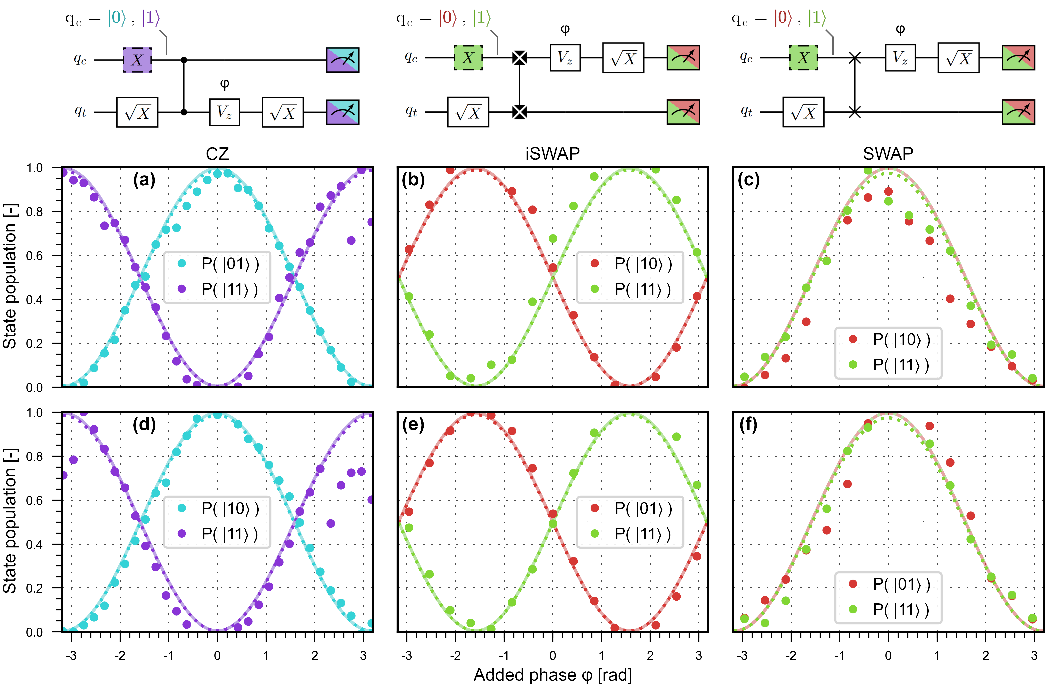}
    \caption{ \scriptsize Ramsey sequences demonstrating the populations and phases resulting from the two-qubit CZ (a, d), iSWAP (b, e), and SWAP (c, f) gates.
    In the upper row of the measured data, the two-qubit gates were performed on the states $\ket{0\,\bar{i}}$ and $\ket{1\,\bar{i}}$, whereas in the lower row, the roles of qubits 1 and 2 were swapped. 
    The data points are post-processed to mitigate state preparation and measurement (SPAM) errors; see \ref{appendix:error_mitigation}. The overlaid sinusoidal curves show the ideal outcome, and the dotted curves take $T_1$ relaxation into account. 
    The CZ, iSWAP, and SWAP results were taken with 35,000 (bottom CZ: 10,000), 15,000 and 250,000 samples per data point respectively. The groups of data that significantly diverge from the ideal curves are discussed further in Section \ref{sec:discussion}. Blue-purple data points relate to conditional Ramsey experiments, while the red-green points relate to cross-Ramsey experiments. The mean squared error between the post-processed data and the ideal curves ranges between 0.001 and 0.020 at an average of 0.007, where the number of samples is 32 for the CZ data and 16 for the iSWAP and SWAP data. Supplementary post-processing data are available in \ref{appendix:error_mitigation}.}
    \label{fig:conditional_and_cross_Ramsey_experiments}
\end{figure}

The transmons are dispersively coupled to coplanar waveguide resonators for state readout, visible in figure~\ref{fig:chip_and_energy_levels}(c, \nolinebreak d). The qubits and couplers are located on one chip, while the resonators and control lines are located on a control-chip interposer \cite{Kosen2022}. The resonators share a common feedline, enabling multiplexed readout of the qubits. The coupler has no dedicated readout resonator.

As our qubits are fixed in frequency, we supply microwave pulses to drive transitions between the energy levels as illustrated in figure~\ref{fig:chip_and_energy_levels}(a). 
The $\ket{0}\to\ket{1}$ transition frequencies of the transmons are $f_1 = 3.860$ GHz and $f_2 = 3.397$ GHz. In these experiments, we bias the coupler with a static magnetic flux at $\Phi = -0.336 \ \Phi_0$ ~\cite{McKay2016}, putting its eigenfrequency at $f_c = 4.863$ GHz, i.e.\@ more than $1$~GHz above both qubits. 
A complete setup description is available in \ref{appendix:experimental_setup}, the system's Hamiltonian in \ref{appendix:energy_dynamics}, and additional system parameters in Table~\ref{tab:parameter_estimation_of_device} of \ref{appendix:tuned_up_parameters}.

Our readout state assignment is configured to distinguish between states $\ket{0}$, $\ket{1}$, and $\ket{2}$. The readout frequency is chosen such that the population spacing in the complex readout plane is maximized. The readout amplitude was then chosen to maximize state-assignment fidelity. The readout duration was not optimized; we selected a readout duration of $2.3$~\textmu s to ensure that we would integrate over the readout trace differences corresponding to the different qubit states \cite{Bianchetti2010_thesis}. 
\ref{appendix:experimental_tuneup} shows additional details of the readout setup, as well as how we used two-qubit confusion matrices to post-process our experimental data \cite{Geller2020, Bialczak2010}.

Our two-qubit gates rely on parametric exchange coupling \cite{McKay2016, Niskanen2006, Reagor2018}, which enables widely detuned fixed-frequency qubits \cite{McKay2016} but with a known drawback of being comparably slow to alternatives such as cross-resonance gates \cite{Kubo2023}. We consider CZ and iSWAP gates as shown in figure \ref{fig:chip_and_energy_levels}(a): to execute iSWAP gates, we resonantly transition between states $\ket{01}$ and $\ket{10}$; to execute CZ gates, we briefly leave the computational subspace by performing a resonant transition between $\ket{11}$ and $\ket{20}$ (and back again) for a tuned-up gate duration $\tau_{\mtext{CZ}}$, completing a round-trip that adds a geometric phase of $\pi$. We perform SWAP operations by compiling Hadamard, $S^\dagger$, iSWAP, and CZ gates, as was derived in Section \ref{sec:SWAP_decomposition_derived}.

\section{Two-qubit gate characterization} \label{section:two_qubit_gate_characterization}

We characterize the two-qubit gate operations in a simplified way in a Ramsey interference experiment, see figure~\ref{fig:conditional_and_cross_Ramsey_experiments}. 
Conventional Ramsey interferometry measures the accrued phase over time $t$ of a single-qubit superposition state, $\ket{0} + e^{i\varphi(t)}\ket{1}$. 
Instead, we use a fixed, short wait time $\tau$ between the $\sqrt{X}$ initialization and analysis pulses, leading to a phase $\varphi_\tau$.
We emulate phase accrual by adding a virtual-Z gate $V_z$ which offsets the phase of each subsequent pulse by $\varphi$.

In the quantum circuits at the top of figure~\ref{fig:conditional_and_cross_Ramsey_experiments}, the upper rail represents the control qubit $q_c$, and the lower rail the target qubit $q_t$. Here, we use notation $\ket{q_c\,q_t}$. We verify that the CZ gate results in the correct qubit populations and phases by using the conditional Ramsey sequence shown in figures \ref{fig:conditional_and_cross_Ramsey_experiments}(a) and \ref{fig:conditional_and_cross_Ramsey_experiments}(d); likewise, we characterize the iSWAP and SWAP operations by using the cross-Ramsey sequence, in figures \ref{fig:conditional_and_cross_Ramsey_experiments}(b), \ref{fig:conditional_and_cross_Ramsey_experiments}(c), \ref{fig:conditional_and_cross_Ramsey_experiments}(e) and \ref{fig:conditional_and_cross_Ramsey_experiments}(f).
In all cases, we first prepare $q_t$ in state $\ket{\bar{i}}$ by using a $\sqrt{X}$ gate, i.e., the Bloch vector is oriented towards $-Y$. We then execute the two-qubit gate of interest, apply the virtual-Z, rotate back by again using a $\sqrt{X}$ gate, and read out in the Z-basis. We verify that the phase of $q_t$ has accrued by the correct amount, depending on the input state $\ket{0}$ or $\ket{1}$ of the control qubit $q_c$.

For all experiments, we sweep the phase $\varphi$ using virtual-Z gates, which determines the Bloch vector longitude to a unique phase coordinate. Consider the CZ gate as an example: the sweep verifies that the state $q_t$ after the CZ gate is $\cos{\frac{\pi}{2}}\ket{0} + e^{i\varphi}\sin{\frac{\pi}{2}}\ket{1}$ and not $\cos{\frac{\pi}{2}}\ket{0} + e^{i(\pi-\varphi)}\sin{\frac{\pi}{2}}\ket{1}$.

\subsection{CZ, by conditional Ramsey sequence}

As seen in equation (\ref{eq:U_CZ}), the conditional-Z gate adds $\pi$ phase onto $q_t$, conditional on $q_c$ being in $\ket{1}$. 
In figure~\ref{fig:conditional_and_cross_Ramsey_experiments}(a) and \ref{fig:conditional_and_cross_Ramsey_experiments}(d), we see that the phase of $q_t$ accrues by $\pi$ if we execute a CZ gate when $q_c$ is in $\ket{1}$. Similarly, we see that no phase accrual occurs if we execute a CZ gate while $q_c$ is in $\ket{0}$. Both of these conditions are necessary for verifying a CZ gate.

\subsection{iSWAP, by cross-Ramsey sequence}
At phase $\varphi = 0$ in figures \ref{fig:conditional_and_cross_Ramsey_experiments}(b) and \ref{fig:conditional_and_cross_Ramsey_experiments}(d), the expected outcome after the iSWAP gate in the cross-Ramsey sequence is the two-qubit state $\ket{+0}$ when $q_c$ is initially prepared in $\ket{0}$, and $\ket{-1}$ when $q_c$ is prepared in $\ket{1}$.
If the virtual-Z gate adds $0$ phase to $q_c$, then the final $\sqrt{X}$ rotation does nothing to the superposition state, resulting in $0.5$ population at state readout. 
As phase $\varphi=\pi/2$ is added, the Ramsey oscillation approaches $\ket{0}$ for $q_c$ initially prepared in $\ket{0}$ but instead assumes $\ket{1}$ for $q_c$ initially prepared in $\ket{1}$.

\subsection{SWAP, by cross-Ramsey sequence}
Just as with the iSWAP case, the SWAP operation in figures \ref{fig:conditional_and_cross_Ramsey_experiments}(c) and \ref{fig:conditional_and_cross_Ramsey_experiments}(f), is verified using a cross-Ramsey sequence. 
The SWAP operation swaps the two states of $q_c$ and $q_t$ with each other, thus $q_c$ is now in state $\ket{\bar{i}}$. The virtual-Z gate adds a phase $\varphi$ onto the state of $q_c$. If $\varphi$ = 0, the final $\sqrt{X}$ gate will rotate $q_c$ to $\ket{1}$, and will otherwise produce $\pi$-periodic Ramsey oscillations between $\ket{0}$ and $\ket{1}$ depending on the phase added. As SWAP gates are non-entangling, any separable input state into the cross-Ramsey sequence involving a SWAP gate will remain separable. One may thus always discuss the single-qubit states throughout the SWAP cross-Ramsey sequence, as was done here.

\section{Discussion and outlook} \label{sec:discussion}

This discussion mainly targets aspects of our post-processing error mitigation and addresses an observed random switching problem with qubit~1.

Figure \ref{fig:conditional_and_cross_Ramsey_experiments} features several clusters of data points that deviate from the ideal outcome. These were acquired close in time.
We further show in \ref{appendix:tuned_up_parameters} that $Q_1$ was afflicted by periods where the acquired Ramsey fringes yielded large fit errors such that $T_2^*$ could not be accurately determined.
The erratic behavior points to the presence of a two-level-system defect strongly coupled to the qubit, with adverse effects on spectral stability and coherence.
Due to the fixed-frequency design of this device's transmons, we do not suspect flux noise significantly contributing to these observed behaviors. However, observing the $T_2^e$ improvement over $T_2^*$ in figure \ref{fig:decoherence_histograms} indicates \cite{Krantz2019} the presence of low-frequency noise.

Our two-qubit gate implementation has relatively long RF pulses, and as such we are interested in a proof-of-principle demonstration of a simplified SWAP decompilation using CZ and iSWAP gates, rather than a complete gate fidelity characterization (which would be performed, e.g., by means of randomized benchmarking).  
Here we estimate the theoretically achievable gate fidelities.
The mean qubit $T_1$ energy relaxation times are 77 \textmu s and 79 \textmu s for $Q_1$ and $Q_2$, respectively, which yield a theoretical~\cite{Abad2022} upper limit of 99.98\% single-qubit fidelity for 20~ns pulses but only 98.8\% two-qubit fidelity for the 890~ns long CZ gate, and 99.2\% for the 640~ns long iSWAP gate.
The total duration of our SWAP cross-Ramsey circuit implementation is 1.960 \textmu s, setting an upper fidelity limit of 97.4\%. The additional duration beyond the CZ and iSWAP gates consists of single-qubit gates and pauses inserted by the compiler. This duration should be compared to the longer 2.67 \textmu s minimum duration required for compiling a SWAP operation using three CZ gates.
Further metrics are available in \ref{appendix:tuned_up_parameters}.

However, we have previously demonstrated parametric two-qubit gates of 250-400 ns \cite{Bengtsson2020, Kosen2022, Warren2023} on comparable quantum processors, and the pulse duration could realistically be further decreased down toward 150~ns, which would enable two-qubit gate errors smaller than 0.1\%.

Furthermore, our SWAP decomposition can be improved by an approach beyond the scope for this paper: the parametric interactions that drive the CZ and iSWAP gates, as shown in figure \ref{fig:chip_and_energy_levels}(a), are in principle executable in parallel. 
More generally, the CZ gate [equation (\ref{eq:U_CZ})] commutes with the iSWAP gate [equation (\ref{eq:U_iSWAP})], and they can be taken in any order in figure~\ref{fig:SWAP_derivation}(d), or even simultaneously.
Tuning up this bichromatic excitation requires frequency adjustments to account for additional dispersive shifts of the coupler and phase compensation offsets \cite{Warren2023}. In practice, realizing such a bichromatic SWAP gate on this specific platform would be possible following upgrades to the underlying instrument control software.
We note that a similar bichromatic gate approach has been suggested and simulated for combining bSWAP and iSWAP gates \cite{Roth2019}.

The SWAP sequence in figure~\ref{fig:SWAP_derivation}(d) may also be simplified by omitting the initial and final Hadamard gates. Considering $U_{\mtext{iSWAP-CZ}}( \Psi ) = $ CZ( iSWAP( $\Psi$ ) ),
\begin{equation} \label{eq:U_iswap_cz}
    U_{\mtext{iSWAP-CZ}}(\Psi) = 
        \left(
        \begin{array}{cccc}
        1 & 0 & 0 & 0 \\
        0 & 0 & i & 0 \\
        0 & i & 0 & 0 \\
        0 & 0 & 0 & -1
        \end{array}
        \right) ~\Psi,
\end{equation}
we may use the $S^{\dagger}$ gates from equation (\ref{eq:U_s_dagger}) on both qubits to convert the off-diagonal $i$ elements into $1$ while also removing the negation on $\ket{11}$,
\begin{equation}
    \left(
    \begin{array}{cc}
    1 & 0 \\
    0 & -i \\
    \end{array}
    \right)^{\otimes 2}
    =
    \left(
    \begin{array}{cccc}
    1 &  0 &  0 &  0 \\
    0 & -i &  0 &  0 \\
    0 &  0 & -i &  0 \\
    0 &  0 &  0 & -1 \\
    \end{array}
    \right)
\end{equation}
\begin{equation} \label{eq:s_dagger_two_on_iSWAP_CZ}
    \left(
    \begin{array}{cc}
    1 & 0 \\
    0 & -i \\
    \end{array}
    \right)^{\otimes 2}
    \left(
    \begin{array}{cccc}
    1 &  0 &  0 &  0 \\
    0 &  0 &  i &  0 \\
    0 &  i &  0 &  0 \\
    0 &  0 &  0 & -1 \\
    \end{array}
    \right)
    =
    \left(
    \begin{array}{cccc}
    1 &  0 &  0 &  0 \\
    0 &  0 &  1 &  0 \\
    0 &  1 &  0 &  0 \\
    0 &  0 &  0 &  1 \\
    \end{array}
    \right),
\end{equation}
in which we see that equation (\ref{eq:s_dagger_two_on_iSWAP_CZ}) is identical to the SWAP unitary matrix from equation (\ref{eq:U_SWAP}), showing that the initial and final Hadamard gates of figure~\ref{fig:SWAP_derivation}(d) may in principle be omitted. Since the conjugate transpose phase gates are implemented as $0$~ns virtual-Z gates, such a SWAP sequence may be as short as the duration of one iSWAP and one CZ gate.

Finally, our current approach to verifying the function of our CZ, iSWAP, and SWAP operations using Ramsey interferometry can be further improved by applying a tomographically complete input set of states \cite{Baldwin2014}. Such a verification uses our current approach as a necessary condition for certifying the CZ, iSWAP, and SWAP operations, while also determining these gate operations to a unique solution of the unitary gate matrices in Section \ref{sec:SWAP_decomposition_derived}.

\section{Conclusion}
We proposed and experimentally confirmed a composite two-qubit SWAP gate compiled from one iSWAP gate followed by one CZ gate. 
We demonstrated this on a subset of our superconducting flip chip-integrated quantum processor using quantum-gate interactions driven by parametric modulation of a coupler. 

This extends the available gate set in this architecture to include the iSWAP gate, reducing the need for three CZ gates to just one iSWAP and one CZ gate for each decompiled SWAP gate.
We characterized the compound SWAP gate using a cross-Ramsey sequence with two-qubit input states, followed by post-processing to account for SPAM errors. 

Going forward, we expect that a sequence of optimized CZ and iSWAP gates with shorter duration can bring the two-qubit gate error down below 0.1\%, even with coherence times at the $100$ \textmu s level. 
We further expect a gate-time reduction by approximately one half by driving both a CZ gate and an iSWAP gate simultaneously.

\ack
For their discussions and assistance, we thank Per Delsing, Miroslav Dob\v{s}\'{i}\v{c}ek, Anita Fadavi Roudsari, G\"{o}ran Johansson, Tong Liu, Simon Pettersson Fors, Hampus Renberg Nilsson, Daryoush Shiri, and Vitaly Shumeiko at Chalmers -- and Riccardo Borgani, Mats O. Thol\'{e}n
at KTH Stockholm.

We acknowledge financial support from the Knut and Alice Wallenberg Foundation (KAW) through the Wallenberg Center for Quantum Technology (WACQT),
and in part from the EU Flagship on Quantum Technology under project HORIZON-CL4-2022-QUANTUM-01-SGA grant 101113946 OpenSuperQPlus100.
A.\@ Frisk Kockum acknowledges support from the Swedish Research Council grant 2019-03696, and the Swedish Foundation for Strategic Research grant FFL21-0279 and grant FUS21-0063.

The devices were fabricated at the Myfab Chalmers nanofabrication laboratory.
Evaporation of bumps and flip-chip bonding was done at the VTT Technical Research center of Finland.

\appendix
\setcounter{section}{0}

\section{Experimental setup} \label{appendix:experimental_setup}

\begin{figure}[ht!]
    \centering
    \includegraphics[width=0.50\textwidth]{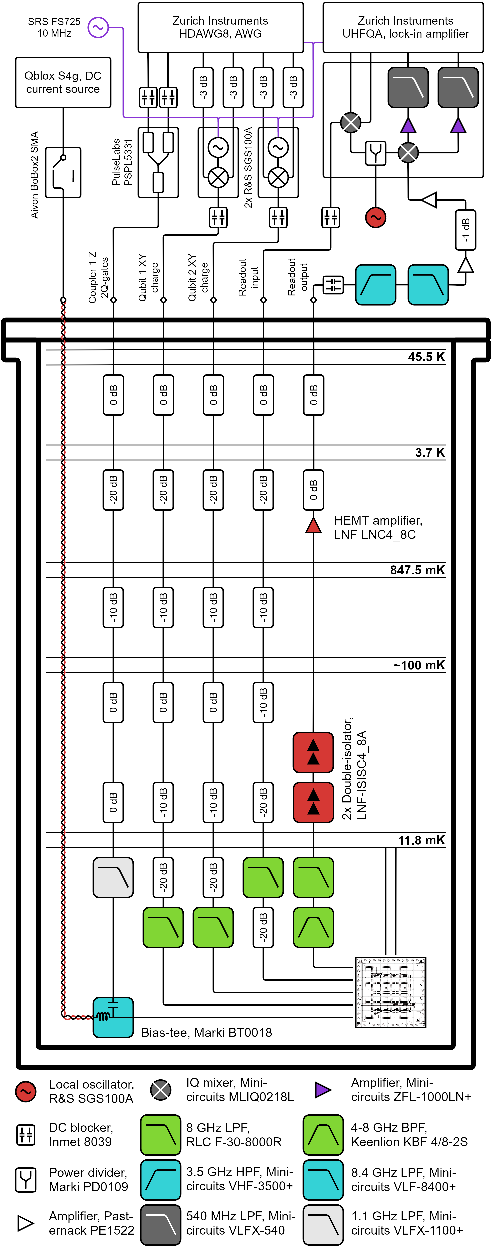}
    \caption{ \scriptsize Our RF setup inside a Bluefors LD250 dilution refrigerator, similar to the suggested layout in \cite{Krinner2019}. LPF, HPF, BPF, denotes low-, high-, and band-pass filters. AWG denotes an arbitrary waveform generator.}
    \label{fig:experimental_setup}
\end{figure}

Figure \ref{fig:experimental_setup} illustrates our cryogenic microwave setup used to perform the experiments in this paper.

\section{Energy dynamics} \label{appendix:energy_dynamics}

Our device shown in figure \ref{fig:chip_and_energy_levels} constitutes a three-body system of multi-level anharmonic oscillators \cite{Koch2007}. The two lowest energy levels of the oscillators define our qubits' computational basis states \cite{Mike_and_Ike}. The anharmonic oscillator in the middle of figure \ref{fig:chip_and_energy_levels}(d) is used to couple together the two qubits on the left and right. We describe our circuit's behavior using the following Hamiltonian:
{\samepage \begin{eqnarray}
\frac{\hat{H}}{\hbar} &=& \sum_{i=1}^{2} \omega_i \hat{a}^{\dagger}_i \hat{a}_i + \frac{\eta_i}{2} \hat{a}^{\dagger}_i \hat{a}_i (\hat{a}^{\dagger}_i \hat{a}_i - 1 ) \\
&& + \omega_c (\Phi) \hat{b}^{\dagger} \hat{b} + \frac{\eta_c}{2} \hat{b}^{\dagger} \hat{b} (\hat{b}^{\dagger} \hat{b} - 1 ) \nonumber \\
&& + \sum_{i=1}^{2} g_{ic} (\hat{a}_i^\dagger + \hat{a}_i) (\hat{b}^\dagger + \hat{b}) \nonumber
\end{eqnarray}}

\noindent
where $\hat{a}$, $\hat{b}$, $\hat{a}^\dagger$, $\hat{b}^\dagger$ are annihilation and creation operators \cite{Gerry_Knight_2004}, with $\hat{a}_i$ acting on transmon $i$ and $\hat{b}$ acting on the coupler; $g_{ic}$ are the coupling strengths between qubits $i$ and the coupler $c$; $\omega_i$ are the angular frequencies of all three transmons, where the coupler transmon's frequency is adjustable by a threading magnetostatic flux $\Phi$; $\eta_i$ and $\eta_c$ are the transmons' anharmonicities \cite{Koch2007}, with suffix $c$ denoting the coupler in particular; and $\hbar = \frac{h}{2\pi}$ is the reduced Planck constant. We assume that the direct qubit-qubit coupling $g_{12} \ll (g_{1,c} , g_{2,c})$. We follow notation $\omega_i = 2 \pi f_i$, where $\omega$ denotes an angular frequency and $f$ the ordinary frequency. 

\section{Parameter estimation} \label{appendix:tuned_up_parameters}

\begin{figure}[t]
    \centering
    \includegraphics[width=0.50\textwidth]{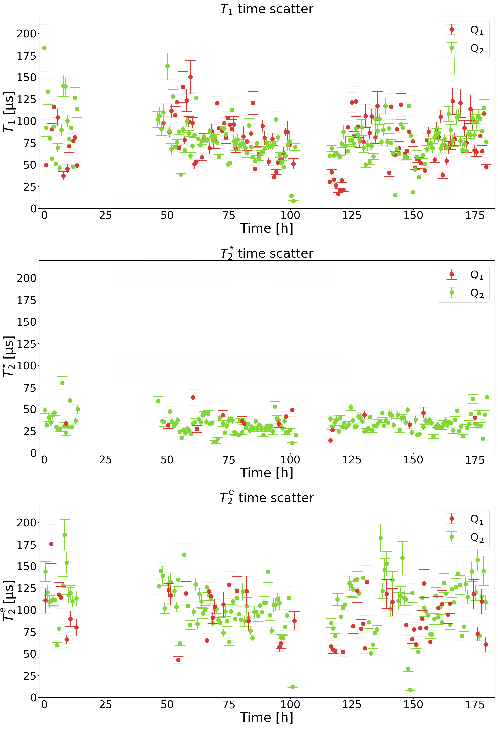}
    \caption{ \scriptsize Time scatter data of decoherence parameters taken during 180 hours. The lack of many fit points for the $T_2^*$ data of $Q_1$ appears to capture this qubit's temporal erratic behavior, keeping in mind that each data point required about 10~min of measurement time.
    Sudden jumps in the data cause fit failures or error bars of more than 15\%, which we enforce as a hard limit on acceptable fit error.}
    \label{fig:decoherence_data}
\end{figure}

\begin{figure}[t]
    \centering
    \includegraphics[width=0.50\textwidth]{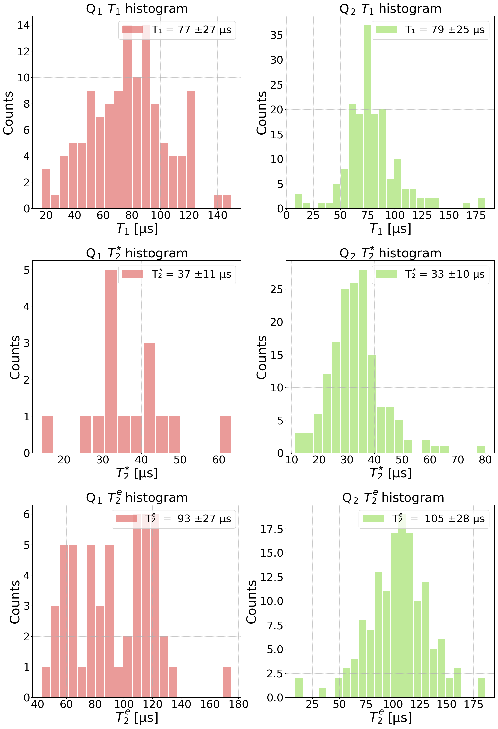}
    \caption{ \scriptsize Histogram distribution of decoherence parameters taken during 180 hours. The number of bins $K$ in the histogram were selected according to the work by Doane \cite{Doane1976}, $K = 1 + \mtext{log}_2( N ) + |\gamma|/\sigma_{\gamma}$, where $N$ is the number of samples, $\gamma$ is the third standardized moment, and $\sigma_{\gamma}$ is the standard deviation of $\gamma$. This approach attempts to highlight skew in a normally distributed dataset while also avoiding oversmoothening distributions.}
    \label{fig:decoherence_histograms}
\end{figure}

Figures \ref{fig:decoherence_data} and \ref{fig:decoherence_histograms} show acquired decoherence data used for establishing the mean $T_1$, $T_2^*$, and $T_2^e$ metrics.
A hard upper limit of 15\% fit error was enforced for all fits;  data where a fit could not be determined with an accuracy below 15\% of the fitted decoherence time $T$ are marked as unreliable and discarded. The fit error is determined from the square root of the diagonal elements of the covariance matrix of the fitted parameters, where the fit equations are

{\samepage \begin{equation}
    Y_{T_1} = A \, e^{{-t} / {T_1}} + m ,
\end{equation}
\begin{equation}
    Y_{T_2^*} = A \, e^{{-t} / {T_2^*}}  \cos{(2 \pi  f  t + \varphi)} + m ,
\end{equation}
\begin{equation}
    Y_{T_2^e} = A \, e^{{-t} / {T_2^e}} + m ,
\end{equation}
}

\noindent where $A$ is an amplitude scalar; $t$ is the time axis stepped through in the decoherence time measurement; $T_1$, $T_2^*$, $T_2^e$ are decoherence times metrics; $f$ is the oscillation rate of the Ramsey fringe; $\varphi$ its apparent phase offset; and $m$ its demodulated magnitude offset.

Table \ref{tab:parameter_estimation_of_device} lists additional parameter data for our device.

\begin{table}
    \footnotesize
    \centering
    \caption{ \scriptsize Physical parameters of our system. The DRAG \cite{Motzoi2009} parameter $\alpha$ is defined using notation from \cite[equation (3.12)]{Bengtsson2020_thesis}. The error bars reported for the frequency offsets $\Delta$
    were calculated using simplified error propagation~\cite{Ku1966}, $\mtext{Err(}\Delta f_{\mtext{CZ}} \mtext{)} = \sqrt{ (-1)^2 \cdot \mtext{Err(} f_{12,\mtext{Q}1} \mtext{)}^2 + (+1)^2 \cdot \mtext{Err(} f_{01,\mtext{Q}2} \mtext{)}^2 + (-1)^2 \cdot \mtext{Err(} f_{\mtext{CZ}} \mtext{)}^2}$,
    $\mtext{Err(}\Delta f_{\mtext{iSWAP}} \mtext{)} = \sqrt{ (+1)^2 \cdot \mtext{Err(} f_{01,\mtext{Q}1} \mtext{)}^2 + (+1)^2 \cdot \mtext{Err(} f_{01,\mtext{Q}2} \mtext{)}^2 + (-1)^2 \cdot \mtext{Err(} f_{\mtext{iSWAP}} \mtext{)}^2}$.
    \newline The error reported for the three-state mean assignment fidelity, $F_{\mtext{Assign\{\left| 0\right>, \left| 1\right>, \left| 2\right>\}}}$, is the standard error from the mean (S.E.M. $= \sigma/\sqrt{3}$), and the variance is $\sigma^2 = \left(\Sigma_{i = 1}^3 (F_{\ket{i}} - F_{\mtext{Assign\{\left| 0\right>, \left| 1\right>, \left| 2\right>\}}})^2\right)/~(3 - 1)$.
    \label{tab:parameter_estimation_of_device}
    }
    \lineup
    \begin{tabular}{cccc}
        \br
        Parameter & $Q_1$, $R_1$ & $Q_2$, $R_2$ & $C_1$ \\
        \mr
        $f_{\mtext{RO},\ket{0}}$ & $6.48084$ GHz & $6.25968$ GHz & n/a \\
        $f_{\mtext{RO},\ket{1}}$ & $6.48068$ GHz & $6.25946$ GHz & n/a \\
        $f_{\mtext{RO},\ket{2}}$ & $6.48056$ GHz & $6.25920$ GHz & n/a \\
        $f_{\mtext{RO},\mtext{optimal}}$ & $6.48048$ GHz & $6.25960$ GHz & n/a \\
        $\chi_{\ket{0} \rightarrow \ket{1}}$ & $80$ kHz & $110$ kHz & \\
        $\chi_{\ket{1} \rightarrow \ket{2}}$ & $60$ kHz & $130$ kHz & \\
        $\tau_{RO}$ & \multicolumn{2}{c}{2.3 \textmu s} & n/a \\
        $F_{\mtext{Assign\{\left| 0\right>, \left| 1\right>, \left| 2\right>\}}}$ & $0.8793$ & $0.8873$ & n/a \\
        & $\pm0.0494$ & $\pm0.0528$ & \\
        \hline
        
        $f_{\ket{0} \rightarrow \ket{1}}$ & $3.86011$ GHz & $3.39741$ GHz & $4.863$ GHz \\
        & $\pm 156$ kHz & $\pm 156$ kHz & \\
        $f_{\ket{1} \rightarrow \ket{2}}$ & $3.60404$ GHz & $3.18992$ GHz & \\
        & $\pm 190$ kHz & $\pm 190$ kHz & \\
        $\eta~/~2\pi$ & $-256.07$ MHz & $-207.49$ MHz & \\
        & $\pm 346$ kHz & $\pm 346$ kHz & \\
        $\tau_{\mtext{ 1QB }}$ & \multicolumn{2}{c}{$20$ ns} & \\
        
        $\alpha_{\mtext{Motzoi}}$ & $-770.5 \cdot 10^{-3}$ & $-221.4 \cdot 10^{-3} $ & \\
        & $\pm13 \cdot 10^{-3}$ & $\pm13 \cdot 10^{-3}$ & \\
        \hline

        $T_{1}$ & $77$ \textmu s $\pm 27$ \textmu s & $79$ \textmu s $\pm 25$ \textmu s & \\
        $T_{2}^*$ & $37$ \textmu s $\pm 10$ \textmu s & $33$ \textmu s $\pm 10$ \textmu s & \\
        $T_{2}^e$ & $93$ \textmu s $\pm$27 \textmu s & 105 \textmu s $\pm 28$ \textmu s & \\
        \hline
        
        $f_{\mtext{LO},XY}$ & \multicolumn{2}{c}{$3.600000000$ GHz} & n/a \\

        $f_{\mtext{LO},RO}$ & \multicolumn{2}{c}{$6.600000000$ GHz} & n/a \\
        \hline
        $\Phi_{\mtext{bias}}$ & \multicolumn{2}{c}{n/a} & $-0.336$ $\Phi_0$ \\
        & \multicolumn{2}{c}{} & $\pm 0.003$ $\Phi_0$ \\
        $f_{\ket{0} \rightarrow \ket{1},~\Phi=0 }$ & $3.8610$ GHz & $3.3977$ GHz & $6.9373$ GHz\\
        $g_{q,c}~/~2\pi$ & 37 MHz & 35 MHz & \\
        \hline
        \hline
        
        \multicolumn{4}{c}{CZ} \\
        \hline
        $\tau_{\mtext{CZ}}$ & \multicolumn{3}{c}{$890$ ns} \\
        $f_{\mtext{CZ}}$ & \multicolumn{3}{c}{$207.24$ MHz $\pm 10 $ kHz}\\
        $\Delta f_{\mtext{CZ}}$ & \multicolumn{3}{c}{$-0.61$ MHz $\pm 246$ kHz}\\
        $\varphi_{\mtext{comp.}}$ & $\mtext{+}0.61$ rad & $\mtext{+}1.03$ rad & $0$ rad \\
        \hline
        \hline
        \multicolumn{4}{c}{iSWAP} \\
        \hline
        $\tau_{\mtext{iSWAP}}$ & \multicolumn{3}{c}{$640$ ns} \\
        $f_{\mtext{iSWAP}}$ & \multicolumn{3}{c}{$461.51$ MHz $\pm 82$ kHz}\\
        $\Delta f_{\mtext{iSWAP}}$ & \multicolumn{3}{c}{$\mtext{+}1.19$ MHz $\pm 235$ kHz}\\
        $\varphi_{\mtext{comp.}}$ & $\mtext{+}2.24$ rad & $\mtext{+}3.36$ rad & $\mtext{+}4.86$ rad \\
        \hline
        \hline
        \multicolumn{4}{c}{SWAP} \\        
        \hline
        $\tau_{\mtext{SWAP}}$ & \multicolumn{3}{c}{1812 ns} \\
        $\varphi_{\mtext{comp.}}$ & $-1.07$ rad & $\mtext{+}1.25$ rad  & $= \varphi_{\mtext{iSWAP}}$ \\
        \br
        
    \end{tabular}
    
\end{table}

\section{Experimental tune-up procedure} \label{appendix:experimental_tuneup}

Our single-qubit gate tune-up mainly follows well-established parameter estimation methods used for superconducting quantum computing \cite{Bengtsson2020_thesis, Chen2018_thesis}.

Our microwave drive pulses are generated using a Zurich Instruments HDAWG8 arbitrary waveform generator (AWG) operating in 300 MHz mode, 16-bit 2.4 GSa/s. Our single-qubit gates are realized as single sideband (SSB) pulses upconverted in frequency using standalone phase-locked mixers with individual local oscillators (Rohde \& Schwarz SGS100A), see the setup shown in figure \ref{fig:experimental_setup}. Note that the coupler tone is not supplied onto the QPU using frequency mixing; the two-qubit gate pulses are generated at baseband on the AWG.

All of our single-qubit operations use a 20 ns Gaussian envelope, whereas for all two-qubit operations this envelope is a flat rectangle with a Gaussian-shaped 10 ns rise time and 10 ns fall time.

Our single-qubit pulses use the Derivative Removal by Adiabatic Gate (DRAG) technique to lessen single-qubit gate coupling to leakage levels \cite{Motzoi2009}. We use DRAG notation from \cite[equation (3.12)]{Bengtsson2020_thesis}, in which the Motzoi parameter is given as $-\alpha/\eta$, where $\alpha$ is a scaling parameter that is tuned-up, and $\eta$ is the transmon anharmonicity in units of angular frequency.

\subsection{Selection of repetition rates and local oscillator frequencies}

When using an AWG to generate microwave pulses at baseband, and then using frequency mixers to upconvert the pulse carrier into the final qubit frequencies as done in figure \ref{fig:experimental_setup}, we must ensure that the repetition rate of every iteration of the experiment is commensurate with the frequency of the mixers' local oscillators as we describe in \cite{Warren2023}. Otherwise, the LO will erroneously gather phase between quantum circuit iterations. This phase error refers to the start of the quantum circuit that is currently being played, which is seen as the reference from which our quantum circuit phases are referred to.

In a scenario where a repetition rate is selected such that it is not phase commensurate with the LO frequency, the CZ and iSWAP tune-up steps where local phase adjustment is determined \cite[equations (C3) and (C11)]{Ganzhorn2020} will still provide measurement results that appear normal for a successful experiment, and a local phase compensation parameter can be extracted from the data. However, for quantum circuits that depend on the phase of the coupler pulse, as is the case when using iSWAP gates, the readout will imply experiment failure by showing maximally mixed states. This erroneous result happens whenever the coupler phase impacts the final state; the phase of the coupler pulse appear to be random, and this random phase information will get lost in the averaging done between readout shots.

We selected an LO frequency that is a multiple of the inverse of the repetition rate, thus being phase commensurate between measurement iterations \cite{Warren2023}. We chose 400 \textmu s, sharing least common multiple 2,500 with $f_{\mtext{LO}} = 3.6$ GHz. At a 400 \textmu s repetition rate, the qubits would on average be expected to naturally relax to $\exp{(-400/78)} = 0.6\%$ population between experiment iterations, which is about three times lower than the residual thermal population of the dilution refrigerator.

\subsection{Coupler biasing} \label{subsection:coupler_biasing_in_appendix}
The coupler is based on a superconducting quantum interference device (SQUID) \cite{Jaklevic1964, Clarke2004}. A SQUID-based coupler allows us to adjust its transmon frequency by threading its central loop with magnetostatic flux, which we generate from a DC bias. This bias is current-controlled using a Qblox S4g 18-bit current source. Using current-controlled sources compensates for variations in resistance of the feedline, for instance due to temperature or moisture dependence of the room-temperature cabling.
Using voltage-controlled sources for SQUID biasing does not stabilize the magnetostatic flux against changes in feedline resistance, thus such errors propagate onto the SQUID biasing.

\begin{figure}[ht!]
    \centering
    \includegraphics[width=0.50\textwidth]{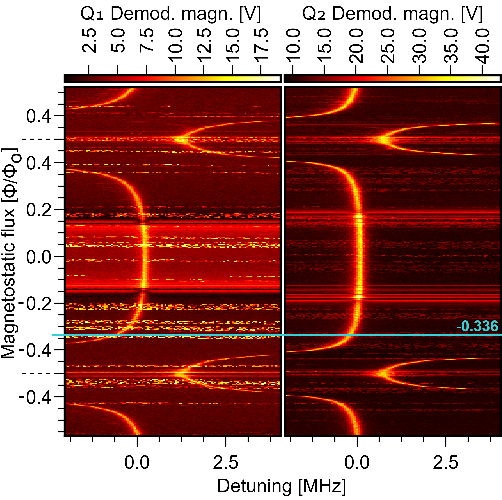}
    \caption{ \scriptsize Two-tone spectroscopy on both qubits, where the coupler bias current ranges from $-1717$ \textmu A to +$3217$ \textmu A. This bias current axis is given as a ratio of one flux-quantum periodicity of the SQUID, calculated using \cite[(2)]{McKay2016}. The marker indicates our chosen bias current at $-0.336 \ \Phi_0$. Note the intermittent noise seen on $Q_1$, to some degree also seen on $Q_2$. Sweeping the bias axis takes $798.53$ min, inferring noise presence in sessions between 2 and 23 min. The horizontal lines spanning the whole spectrum at $\pm 0.16 \ \Phi_0$ and $\pm 0.20 \ \Phi_0$ show where the coupler's frequency tunes past the resonator frequencies.}
    \label{fig:coupler_bias_sweep}
\end{figure}

Since our device does not feature a resonator for reading out the SQUID coupler, we instead perform SQUID parameter estimation via its coupled qubits, specifically two-tone spectroscopy as a function of the applied magnetostatic coupler bias as is shown in figure \ref{fig:coupler_bias_sweep}.

We initially select a magnetostatic bias close to $0.5 \ \Phi_0$ for our tunable coupler, which is nominally identical to $-0.5 \ \Phi_0$ (see figure \ref{fig:coupler_bias_sweep} and \cite[equation (2)]{McKay2016}). Next, we proceed with setting up readout state assignment, as will be explained in \ref{appendix:readout_tune_up}. We then proceed with two-qubit gate tune-up. Since the readout state assignment has been tuned up, we may fine-tune the coupler flux bias to prevent leakages into undesired states while modulating the coupler with the carrier tone of a two-qubit gate. To find such a flux bias, we first prepare state $\ket{11}$. Then, we sweep the coupler's $\{\omega_c, \Phi\}$ parameter space while applying a constant carrier tone onto the coupler that equals the $\ket{11} \rightarrow \ket{20}$ transition frequency. We avoid selecting biases where we discern that this modulation induces population transfer into unwanted states. As can be seen in figure \ref{fig:coupler_bias_sweep}, the qubits' transition frequencies is expected to shift on the order of a few MHz as a function of applied flux; it is good practice to be aware of the spectral width of the applied $\ket{11} \rightarrow \ket{20}$ coupler tone, to ensure that the $\ket{11} \rightarrow \ket{20}$ transition is still driven as the qubits move in frequency.

\subsection{Readout setup and optimization} \label{appendix:readout_tune_up}

We perform state readout using a Zurich Instruments UHFQA lock-in amplifier, 600 MHz, 12/14-bit ADC/DAC 1.8 GSa/s, facilitated by a combination of microwave devices for frequency conversion as is shown in figure \ref{fig:experimental_setup}. The readout is performed in transmission mode where the $S_{21}$ transmission function is used for readout parameter estimation, as shown in figure \ref{fig:resonator_three_traces}.

We set up the lock-in amplifier to discard the first 1,000 scoped samples, to allow for signal propagation through the setup and resonator ring-up. We collect the following 4,096 samples, corresponding to 2.3 \textmu s, which makes up two time traces corresponding to the in-phase and quadrature (IQ) components of the signal. These two time traces are demodulated to baseband, and integrated, which forms a single complex data point in the IQ plane. After averaging, each data point in figure \ref{fig:resonator_three_traces} corresponds to such a data point.

\subsubsection{Frequency optimization for state assignment}
We proceed with optimizing the frequency of the resonator ring-up tone used for state readout, which we denote as the ``readout frequency''. This frequency optimization is performed by sweeping the readout tone's carrier frequency while monitoring magnitude and phase of the ensuing three frequency responses for $\{\ket{0},\ket{1},\ket{2}\}$. To minimize data transfer between the PC and the lock-in amplifier, changing the readout tone sent onto the resonators is done by stepping the LO frequency of the mixer connected to the lock-in amplifier's pilot tone.

\begin{figure}[ht!]
    \centering
    \includegraphics[width=0.50\textwidth]{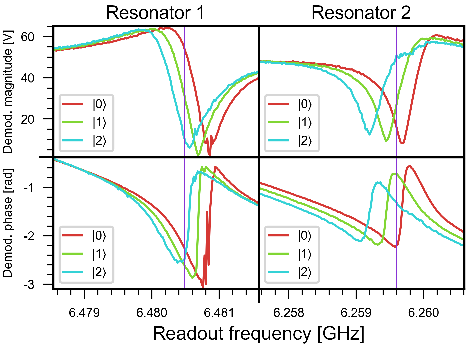}
    \caption{ \scriptsize Readout magnitude and phase traces of our two resonators, where the ground, excited, and second-excited states of the qubits show visible impact on the frequency and phase transfer functions. The dispersive shifts are $\chi_{\ket{0}\to\ket{1}} = $ 80 kHz and 110 kHz, and $\chi_{\ket{1}\to\ket{2}} = $ 60 kHz and 130 kHz, on resonators 1 and 2, respectively. The vertical lines indicate the readout frequencies at which the magnitude-phase separation on the complex plane was maximized for the two resonators' three readout traces. As optimal readout frequencies, 6.48048 GHz and 6.25960 GHz were selected for resonators 1 and 2, respectively.}
    \label{fig:resonator_three_traces}
\end{figure}

Since we have acquired the qubit control pulse parameters for the $\ket{0}\to\ket{1}$ and $\ket{1}\to\ket{2}$ transitions of both qubits, we may place the qubits in the three respective states, and trace out the IQ coordinates for every step along the x-axis of figure \ref{fig:resonator_three_traces}. We analyze the distance between the three respective IQ coordinates for all frequency steps, in order to select an optimal readout frequency. The parameter being optimized for, is the perimeter spanned by the three population centers of the readout states; optimizing for different parameters may better suit certain usage cases, one optimization alternative is to minimize the worst distance between any two states.
It is crucial to note that the resonator frequency responses at this time are acquired using averaged measurements; each data point is the result of repeating the same experiment 1,024 times.

\subsubsection{Amplitude optimization for state assignment}

Using the optimized readout frequency, we instead sweep the IF amplitude of the readout tone, for the three considered states of the two resonators.
It is crucial to note that the previous averaging of 1,024 is now set to 1, i.e., we perform single-shot readouts from this point onward.

\begin{figure}[t!]
    \centering
    \includegraphics[width=1.00\textwidth]{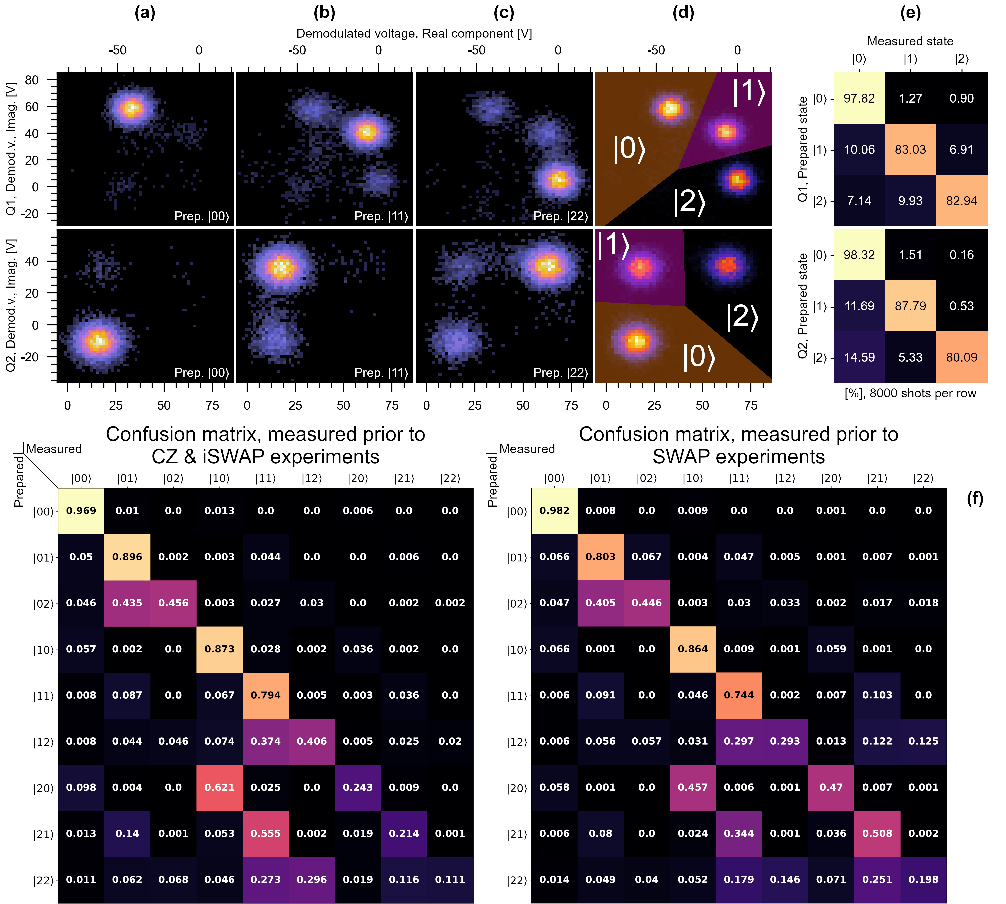}
    \caption{ \scriptsize Readout state assignment setup. The top row shows the setup for qubit 1 and the second row for qubit 2. Columns (a)--(c) show the analogue demodulation of the ground, excited, and second-excited states, respectively. Column (d) shows state assignment plots given the population distribution. Column (e) shows the single-qubit confusion matrices. State assignment is done by assigning a single-shot measurement to its nearest population center of the three considered states. We emphasize that the collected data is single-shot, and not averaged, before being assigned to a state. The two top rows were acquired using 8,000 shots per subplot. The single-qubit confusion matrices yield assignment fidelities of 87.93\% $\pm$4.95\% and 88.73\% $\pm$5.28\%, which is comparable to the two-qubit planar reference design \cite{Bengtsson2020} on which this device is based, and 5-6\% less than the uncorrected readout fidelity of our more recent devices \cite{Chen2023}. We observe 7-15\% classification error into $\ket{0}$: considering $T_1 \approx 77$ \textmu s, the 2.3 \textmu s readout pulse is expected to yield $\approx$ 3\% energy relaxation error into $\ket{0}$ during the readout; the thermal population error is estimated to $|\braket{1|\Psi_0}|^2 + |\braket{2|\Psi_0}|^2 \approx$ 2\%; the remaining 2-10\% error in $\ket{0}$ is expected to stem from gate error, overlap error, and readout-induced mixing as DRAG is used. Two separate multi-qubit-state confusion matrices were acquired for the conditional and cross-Ramsey experiments in figure \ref{fig:conditional_and_cross_Ramsey_experiments}, corresponding to two different measurement sessions. These sessions experimented on either the CZ gate and iSWAP gate [row (f), left matrix], or the SWAP gate [row (f), right matrix]. These two-qubit confusion matrices were acquired using 25,000 (CZ, iSWAP) and 120,000 (SWAP) shots. Figure \ref{fig:confusion_matrix_difference} compares two-qubit confusion matrix data to the single-qubit confusion matrices, showing that these two-qubit confusion matrix data are overall very similar, differing mainly in $\ket{2} \rightarrow \ket{1}$ readout errors.}
    \label{fig:readout_error_mitigation}
\end{figure}

The two qubits are prepared in the ground state by natural relaxation, meaning no state preparation is performed. We then probe the resonators at their optimized frequency 8,000 times, with averaging disabled. The result of these 8,000 state readouts form a 2D histogram, showing a distinct population in the complex space, which can be seen in figure \ref{fig:readout_error_mitigation}(a). This experiment is repeated, but preparing the qubits in the excited state. The resonator readout 2D histograms now show a population that has moved in the complex space, as is visible in figure \ref{fig:readout_error_mitigation}(b). Similarly, this experiment is repeated for the second-excited state, seen in figure \ref{fig:readout_error_mitigation}(c).

The three populations all feature a population center: a coordinate in the complex plane that corresponds to the average of the 8,000 shots performed in order to create that population in the histogram. We now assign each one of the $3 \cdot \mtext{8,000}$ shots to its closest population center, which creates three distinct zones in the complex plane where any shot that ends up in a certain zone, is assigned a distinct discrete state. Both resonators feature three such zones, seen in figure \ref{fig:readout_error_mitigation}(d).

Since we know which was the intended state for all ${3 \cdot \mtext{8,000}}$ shots of this experiment using the optimized readout frequency at some amplitude, we may now also count how many shots missed their target, i.e., how many shots are not assigned to their prepared (intended) state. The results of this counting is shown in figure \ref{fig:readout_error_mitigation}(e). The trace of these matrices is commonly known as the assignment fidelity. Our assignment fidelity is defined from the three states we consider in our readout,

\begin{equation} \label{eq:assignment_fidelity}
    F_{\mtext{Assignment,}R_j} = \frac{|\braket{0|\Psi_{0}}|^2 + |\braket{1|\Psi_{1}}|^2 + |\braket{2|\Psi_{2}}|^2}{3}
\end{equation}

\noindent
where $\Psi_{i}$ is the state of the qubit that corresponds to resonator $R_j$ when this qubit was prepared in state $\ket{i}$.

\begin{figure}[ht!]
    \centering
    \includegraphics[width=0.50\textwidth]{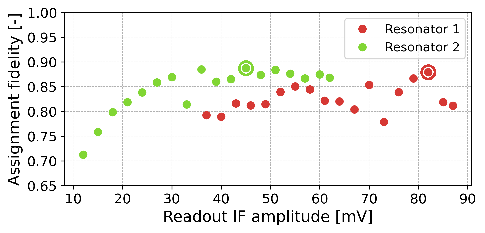}
    \caption{ \scriptsize State-assignment fidelity calculated from equation (\ref{eq:assignment_fidelity}) as a function of the readout tone IF amplitude. The amplitudes for the two resonators' readout pulses are selected at the highest assignment fidelity, which were 0.8793 $\pm$4.95\% and 0.8873 $\pm$5.28\% for $Q_1$ and $Q_2$ respectively.}
    \label{fig:readout_amplitude_vs_assignment_fidelity}
\end{figure}

The state-assignment process is repeated for a sweep of 18 different IF amplitudes, ranging from $37$ mV to $87$ mV for resonator 1, and from $12$ mV to $62$ mV for resonator 2. We select the readout IF amplitudes which give us the highest assignment fidelities when reading out the two resonators, $82$ mV and $45$ mV, shown in figure \ref{fig:readout_amplitude_vs_assignment_fidelity}. All plots in figure \ref{fig:readout_error_mitigation} were taken at the optimized frequency and optimized amplitude. We employ no further readout optimization.

From this point onward, whenever we perform single-shot measurements, each readout yields one data point in the complex plane. We assign this data point to its closest population center, see figure \ref{fig:readout_error_mitigation}(d).

We do not employ matched filtering techniques \cite{Tholen2022} where the demodulated signal is compared to known time traces corresponding to the resonators' frequency responses for the ground, excited, or second excited states of the qubits.

\subsection{Two-qubit gate tune-up} \label{appendix:two_qubit_gate_tune_up}

At this point of the tune-up process, all readout and single-qubit parameters are known, and the DC current bias for the coupler is set. To perform two-qubit gates, we modulate the coupler with a microwave tone giving rise to the magnetic flux

\begin{equation} \label{eq:coupler_ac}
    \Phi_c(t) = A(t) \cdot \cos{( 2\pi f_c t + \varphi)} + \Phi_{\mtext{DC}}
\end{equation}

\noindent
where $f_c$ is the coupler pulse's carrier frequency, $\varphi$ is the coupler pulse's phase relative to the start of the quantum circuit, $A(t)$ is the pulse envelope, and $\Phi_{\mtext{DC}}$ is the static offset bias which sets the unmodulated frequency of the coupler.

\ref{subsection:coupler_biasing_in_appendix} explained how $\Phi_{\mtext{DC}}$ was set using a current-controlled bias from a current generator. Parameters $A$, $f_c$, $\varphi$ are properties of the IF tone generated by the AWG. As can be seen in figure \ref{fig:experimental_setup}, the AC component of equation (\ref{eq:coupler_ac}) is combined with the bias current at the mixing chamber of the dilution refrigerator.

When performing two-qubit gate tune-up, we are often forced to repeat a subset of the flux bias tune-up of the coupler; we would initially select some coupler bias, then engage in CZ and iSWAP tune-up, knowing that we might have to adjust the coupler bias. Changing the coupler bias requires re-tuning the readout assignment parameter. New state assignment parameters are often similar to those found previously; spectrally, the coupler's transition frequency is typically located $\sim$1400 MHz away from the closest considered resonator transition, meaning little change to the readout parameters upon adjustment.

One example of a failure point could be that for some selected CZ amplitude and carrier frequency, population would transfer from $\ket{11} \rightarrow \ket{20}$ at $\frac{\tau_{\mtext{CZ}}}{2}$. But, as $\tau_{\mtext{CZ}}$ is swept to determine the CZ gate time, we would not see characteristic population oscillations; instead, we might see a single localized area where population transfers from ${\ket{11} \rightarrow \ket{20}}$ but does not come back to $\ket{11}$ at time $\tau_{\mtext{CZ}}$. Another example could be that CZ gates can be tuned up, but iSWAP gates would show close to no swapping ability between $\ket{10} \leftrightarrow \ket{01}$.

\subsubsection{Parameter estimation}

For tuning up CZ and iSWAP gates, our tune-up follows the procedure outlined by Ganzhorn \textit{et al.} \cite{Ganzhorn2020}. We perform conditional Ramsey and cross-Ramsey experiments of these gates to verify their behavior, and ensure that the gates are phase-corrected as can be seen in figure \ref{fig:conditional_and_cross_Ramsey_experiments}.

Our iSWAP drive frequency of $461.51$ MHz, e.g., the pulse carrier, lies outside of the $300$ MHz baseband bandwidth of the AWG, which results in a significant roll-off in amplitude output from the instrument as compared to what was defined in software. This roll-off does not prevent the iSWAP gate from being tuned up as the AWG is capable of making up for the amplitude lost to roll-off.

In superconducting quantum processor architectures that use fixed-frequency qubits with tunable couplers, local qubit phase adjustments are required after applying two-qubit gates, as the coupler drive pulse itself is dispersively shifting the coupler \cite{Ganzhorn2020}, consequently changing the qubit frequencies as was seen in figure \ref{fig:coupler_bias_sweep}. Using virtual-Z gates, we adjust for these local phase offsets of the iSWAP and CZ gates, in a similar manner to \cite{Ganzhorn2020}.

As is shown in \cite{Abrams2020}, using virtual-Z gates for local phase adjustment requires adjusting the coupler drive for the iSWAP gates with $\varphi_c = \varphi_{2} - \varphi_{1}$, where $\varphi_{1}$ and $\varphi_{2}$ are the local qubit phase adjustments done for the iSWAP gate. We do not adjust the phase of the coupler drive for CZ gates, as the CZ gate is invariant to the phase of the coupler pulse \cite{Ganzhorn2020, Reagor2018}.

\begin{figure}[ht!]
    \centering
    \includegraphics[width=0.50\textwidth]{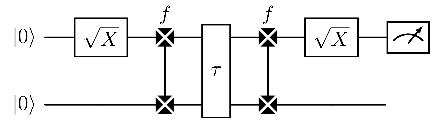}
    \caption{ \scriptsize Experiment for measuring the two-qubit frame $f_{\mtext{2qf}}$. The gate element shown as $\tau$ is a time delay between the two iSWAP gates. Begin the experiment by assuming some coupler drive frequency $f$ close to the tuned-up $f_{\mtext{iSWAP}}$. As $\tau$ increases, the readout will show a Ramsey-like oscillation dependent on the detuning between $f$ and $f_{\mtext{2qf}}$. Once $f$ matches the frequency of the two-qubit frame $f_{\mtext{2qf}}$, the oscillation frequency of the Ramsey-like readout will be 0 Hz. Therefore, a two-dimensional sweep of $f$ and $\tau$ reveals $f_{\mtext{2qf}}$ graphically, similar to a two-dimensional Ramsey interferometry.}
    \label{fig:two_qubit_frame_measurement}
\end{figure}

In addition to this coupler phase offset, the coupler drive must be further phase adjusted to account for frame tracking, as the iSWAP gate is not resonant with the qubit frames. The ideal frequency $f_{\mtext{2qf}}$ at which an iSWAP gate is performed, equals the frequency corresponding to the energy difference between $|\ket{10} - \ket{01}|$. However, as the qubits move in frequency due to the coupler being affected during the coupler drive pulse playing, the tuned-up iSWAP gate carrier frequency $f_{\mtext{iSWAP}} \neq f_{\mtext{2qf}}$. This frequency difference between the coupler drive and the two-qubit frame leads to a phase offset accumulating during the idle time between two subsequent iSWAP gates, which must be compensated for. To experimentally measure $f_{\mtext{2qf}}$ for iSWAP gates, consider the quantum circuit in figure \ref{fig:two_qubit_frame_measurement}.

\begin{figure}[ht!]
    \centering
    \includegraphics[width=0.50\textwidth]{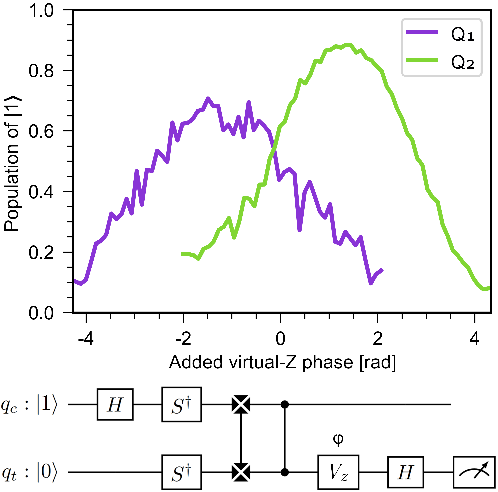}
    \caption{ \scriptsize Parameter estimation of the SWAP local phase correction, shown as the population in $\ket{1}$ on a target qubit $q_t = \{Q_1,Q_2\}$ as the virtual-Z phase $\varphi$ is swept. The control qubit $q_c = \{Q_2,Q_1\}$ is prepared in $\ket{1}$ with an initial $X$ gate. The local qubit phase corrections are selected as -1074.1 mrad and +1253.0 mrad for qubits $Q_1$ and $Q_2$ respectively. The traces shown are not error-corrected for SPAM errors, hence the comparably low population at the two maxima. Each data point was taken using 30,000 single shots, where states \{$\ket{0}$,$\ket{1}$,$\ket{2}$\} were considered in the state assignment. Recall that the $S^\dagger$ gates are implemented as 0 ns virtual-Z gates.}
    \label{fig:SWAP_phase_tuneup}
\end{figure}

In the cross-Ramsey interferometry experiments performed for this work, we are only concerned with the coupler phase of the iSWAP gates at two distinct moments: the first moment is after one initial $\sqrt{X}$ gate has played in the iSWAP cross-Ramsey interferometry, and the second moment is after the Hadamard gate has played after preparing $\ket{q_c q_t} = \ket{0\bar{i}}$, or $\ket{1\bar{i}}$, in the SWAP cross-Ramsey interferometry. In the case of the first moment, the frame-tracked coupler phase of the iSWAP gate was acquired from the iSWAP tune-up procedure. In the second moment, we measured the acquired local phase offset that occurred from playing the same iSWAP gate but at a different time in the quantum circuit, i.e., after executing the first Hadamard of the SWAP sequence in figure~\ref{fig:SWAP_derivation}(d) on a prepared input state. Figure \ref{fig:SWAP_phase_tuneup} demonstrates how the local phase of the qubits was impacted from playing the same iSWAP coupler phase as was done for the iSWAP cross-Ramsey interferometry.

We adjust the local phases after the CZ gate has finished playing with the phase value that that places the investigated qubit as close as possible to $\ket{1}$ in figure \ref{fig:SWAP_phase_tuneup}, as this state is the expected outcome. As such, for our work, we ensure frame-tracked iSWAP gates through the means of local phase updates. However, we emphasize that for a generic implementation using iSWAP gates, the frame tracking could just as well have been done on the coupler drive.

\section{SPAM error mitigation} \label{appendix:error_mitigation}

Let us consider one (any) of the twelve traces of data from figure \ref{fig:conditional_and_cross_Ramsey_experiments}. For each step along the x-axis of an experiment, we measure some data point $\overrightarrow{y}$ that contains the observed probability of all nine considered two-qubit states, ${(\mtext{3~states})}^{(\mtext{2~qubits})}$:

\begin{equation} \label{eq:y_vector}
    \overrightarrow{y} = \left[
|\alpha_{00}|^2, |\alpha_{01}|^2, |\alpha_{02}|^2, |\alpha_{10}|^2, |\alpha_{11}|^2, ... , |\alpha_{22}|^2
\right]^\mtext{T}
\end{equation}

\noindent
where $\alpha_{i}$ is the complex amplitude of the two-qubit state $\ket{i} \in \{0,1,2\}^2$ of the system's wave function $\Psi$ at measurement.

Before the conditional and cross-Ramsey experiments, we intentionally prepared $\Psi$ in all nine two-qubit states, and measured the outcome from thousands of iterations as shown in figure \ref{fig:readout_error_mitigation}. From these empirical statistics, we construct a right stochastic \cite{Szabo2015} matrix:

\begin{equation} \label{eq:00_given_00etc}
    T = \left[ \begin{array}{cccc}
    |\braket{00|\Psi_{00}}|^2 & |\braket{01|\Psi_{00}}|^2 & \cdots & |\braket{22|\Psi_{00}}|^2\\
    |\braket{00|\Psi_{01}}|^2 & |\braket{01|\Psi_{01}}|^2 &  & \\
    \vdots & & \ddots & \\
    |\braket{00|\Psi_{22}}|^2 & & & |\braket{22|\Psi_{22}}|^2
    \end{array} \right]
\end{equation}

\noindent
where each row shows the most likely distribution of measurement outcomes from observing the prepared two-qubit state $\Psi_i$, given that some two-qubit state ${\ket{i} \in \{0,1,2\}^2}$ was prepared for that row.

Each element of the matrix shows the conditional probability $P( \ket{j}$ \textit{measured} $\mid \ket{i}$ \textit{prepared} $)$, where ${j,i \in \{0,1,2\}^2}$. Our experimentally acquired values for matrix $T$ in equation (\ref{eq:00_given_00etc}) was shown in the bottom row of figure \ref{fig:readout_error_mitigation}.

We now assume that our acquired data point $\overrightarrow{y}$ is the result of manipulating some ``true value" data point $\overrightarrow{x}$,

\begin{equation} \label{eq:x_vector}
    \overrightarrow{x} = \left[
|\beta_{00}|^2, |\beta_{01}|^2, |\beta_{02}|^2, |\beta_{10}|^2, |\beta_{11}|^2, ... , |\beta_{22}|^2
\right]^\mtext{T}
\end{equation}

\noindent
which was the true outcome of the experiment before $\overrightarrow{x}$ was manipulated by $T$, resulting in the actual measured output $\overrightarrow{y}$.

\begin{equation}
    \overrightarrow{y} = T\overrightarrow{x}
\end{equation}

\begin{figure}[ht!]
    \centering
    \includegraphics[width=0.50\textwidth]{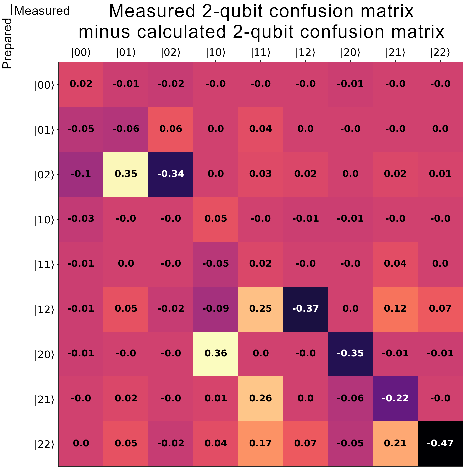}
    \caption{ \scriptsize Matrix showing $C_\mtext{Measured} - C_\mtext{Constructed}$, where $C_\mtext{Measured}$ is the confusion matrix measured prior to the SWAP experiments from figure \ref{fig:readout_error_mitigation}, and $C_\mtext{Constructed}$ is the two-qubit confusion matrix acquired from combining the measured single-qubit confusion matrices from figure \ref{fig:readout_error_mitigation}. This matrix highlights effects that stem from two-qubit readout, such as measurement-induced state decay and readout crosstalk, as seen below the diagonal. From preparing $\ket{12}$, we see that $C_\mtext{Measured}$ yielded 0.19 more population above the diagonal than $C_\mtext{Constructed}$, which we believe hints at measurement-induced transition errors. Apart from $|\braket{12|\Psi_{22}}|^2$, the remaining six combinations where a qubit could decay $\ket{2} \rightarrow \ket{1}$, all accounted for the largest observed population increases: \{0.36, 0.35, 0.26, 0.25, 0.21, 0.17\}.}
    \label{fig:confusion_matrix_difference}
\end{figure}

Remember that for every data point in any given trace from figure \ref{fig:conditional_and_cross_Ramsey_experiments}, we acquired a vector $\overrightarrow{y} = T\overrightarrow{x}$ during the experiment itself. Figure \ref{fig:confusion_matrix_difference} highlights that $T$ must be acquired by preparing and measuring all possible qubit state combinations, and that assembling $T$ from its single-qubit confusion matrices yields different results.

Once we have acquired $T$, we use an implementation \cite{scipy_lsq_linear} of the STIR algorithm \cite{Branch1999} for minimizing the difference

\begin{equation} \label{eq:lm}
    \min \left( \frac{1}{2} \cdot | \overrightarrow{y} - T\overrightarrow{x} |^2 \right) .
\end{equation}

Specifically, since $\overrightarrow{y}$ and $T$ were found empirically and are thus known, the algorithm finds the vector $\overrightarrow{x}$ that minimizes equation (\ref{eq:lm}), meaning that $\overrightarrow{x}$ is now our best guess as to the non-distorted outcome of the experiments that resulted in $\overrightarrow{y}$ when distorted by $T$, for all data points.

Using notation $\ket{q_c\,q_t}$: in figure \ref{fig:conditional_and_cross_Ramsey_experiments}, we plotted the calculated $|\beta_{01}|^2$ and $|\beta_{11}|^2$ from equation (\ref{eq:x_vector}) for the CZ conditional Ramsey traces. And similarly for the iSWAP and SWAP cross-Ramsey traces, we plot $|\beta_{10}|^2$ and $|\beta_{11}|^2$. Remember that the roles of the qubits in the top row of figure \ref{fig:conditional_and_cross_Ramsey_experiments} is $\ket{q_c\,q_t} = \ket{Q_2\,Q_1}$ for the CZ experiments, and $\ket{q_c\,q_t} = \ket{Q_1\,Q_2}$ for the iSWAP and SWAP experiments. In the bottom row of figure \ref{fig:conditional_and_cross_Ramsey_experiments}, the two qubits exchange roles in the quantum circuits.

Table \ref{tab:ramsey_data} shows numerical results of this best-guess reconstructed $|\beta_{01}|^2$ and $|\beta_{11}|^2$ for the CZ experiments, and $|\beta_{10}|^2$ and $|\beta_{11}|^2$ for the iSWAP and SWAP experiments. We note that the mean squared error (MSE) on average is $0.007$ where $0$ corresponds to no error between datapoints and the ideal curve; where, the CZ datasets contained $32$ samples per trace and the iSWAP and SWAP datasets contained $16$ samples per trace. A notable outlier is seen with the $|\beta_{11}|^2$ reconstruction for the CZ trace where $\ket{q_c\,q_t} = \ket{Q_2\,Q_1}$ at MSE $= 0.02033$, which is twice as large as the second largest mean squared error. We attribute this outlier to the two notably deviating clusters of data in figure \ref{fig:conditional_and_cross_Ramsey_experiments}(d), P($\ket{11}$), which were discussed in Section \ref{sec:discussion}.

\begin{table}[ht!]
    \footnotesize
    \centering
    \caption{ \scriptsize Conditional Ramsey and cross-Ramsey data of figure \ref{fig:conditional_and_cross_Ramsey_experiments}. \textit{P, swing} is the fitted population swing where the ideal swing is from $0.0$ to $1.0$, i.e., the expected ideal number is $1.0$; \textit{P,} $\Delta$\textit{offset} is the difference $m_{\mtext{ideal}} - m$ between the expected ideal offset $m_{\mtext{ideal}} = 0.5$ and the fitted offset $m$, where $m_{\mtext{ideal}} = 0.5$ denotes that the expected population swing is oscillating about $0.5$, and \textit{P,} $\Delta$\textit{offset} $= 0.0$ means no difference between the expected ideal and the experimental data; $\Delta$ \textit{phase} is similarly the deviation in the oscillating population's phase, where $\Delta$ \textit{phase} $= 0.0$ means no difference between the expected phase and the experimental data; and MSE denotes the mean squared error: $\mtext{MSE} = \frac{\Sigma(y - p)^2}{N}$, where $y$ is the datapoint, $p$ is its predicted ideal value, and $N$ is the number of datapoints. An MSE of $0$ means no error between the datapoints and the ideal prediction. \label{tab:ramsey_data}
    }
    \lineup
    \begin{tabular}{cccccc}
        \br
        Operation  & Output & P, swing & P, $\Delta$ offset & $\Delta$ phase & MSE \\ 
        $q_c, q_t$ & state  & [-]         & [-]                   & [mrad]      & [-]    \\
        \mr
        CZ         & $\ket{01}$ & 0.987(1) & 0.027(11)  & 37(8)   & 0.00117 \\
        $q_1, q_2$ & $\ket{11}$ & 0.967(3) & 0.077(15)  & \-57(15)  & 0.00632 \\
        \hline
        CZ         & $\ket{10}$ & 1.008(13) & \-0.006(8)  & 71(12)  & 0.00114 \\
        $q_2, q_1$ & $\ket{11}$ & 0.868(41) & 0.150(17) & \-16(52)  & 0.02033 \\
        \hline
        iSWAP      & $\ket{10}$ & 1.079(18) & \-0.023(6)  & 33(16)  & 0.00548 \\
        $q_1, q_2$ & $\ket{11}$ & 1.064(29) & \-0.060(10) & \-85(26)  & 0.00915 \\
        \hline
        iSWAP      & $\ket{01}$ & 1.041(5)  & 0.064(2)  & 85(4)   & 0.00401 \\
        $q_2, q_1$ & $\ket{11}$ & 1.057(9)  & \-0.038(3)  & 75(8)   & 0.00570 \\
        \hline
        SWAP       & $\ket{10}$ & 0.849(33) & 0.049(16) & \-96(42)  & 0.00997 \\
        $q_1, q_2$ & $\ket{11}$ & 0.859(44) & 0.023(27) & \-68(47)  & 0.00584 \\
        \hline
        SWAP       & $\ket{01}$ & 0.954(39) & \-0.055(20) & 223(44) & 0.00953 \\
        $q_2, q_1$ & $\ket{11}$ & 0.993(23) & \-0.050(12) & 103(26) & 0.00327 \\
        \br
    \end{tabular}
\end{table}

\section*{Data and code availability}
The data and code supporting this work has been made available in the Zenodo open repository \cite{Zenodo_data_and_code}.

\section*{References}

\bibliographystyle{iopart-num.bst}
\bibliography{master.bib}

\end{document}